\documentclass{article}
\usepackage[utf8]{inputenc}
\usepackage{authblk}
\usepackage{indentfirst}
\usepackage{hyperref}
\usepackage{longtable}
\usepackage{setspace}
\usepackage[margin=1in]{geometry}
\usepackage{graphicx}
\graphicspath{ {./figures/} }
\usepackage{subcaption}
\usepackage{amsmath}
\usepackage{lineno}
\usepackage{multirow}
\usepackage{color,soul}
\usepackage[version=3]{mhchem}
\usepackage{tabularx}
\usepackage{array}
\newcolumntype{Y}{>{\raggedright\arraybackslash}X} %
\usepackage[biblabel]{cite}

\title{Evaluation of Foundational Machine Learned Interatomic Potentials for Migration Barrier Predictions}

\author[1,2]{Achinthya Krishna Bheemaguli}
\author[3,*]{Penghao Xiao}
\author[2,*]{Gopalakrishnan Sai Gautam}

\affil[1]{Department of Metallurgical and Materials Engineering, National Institute of Technology Karnataka, Surathkal 575025, India}
\affil[2]{Department of Materials Engineering, Indian Institute of Science, Bengaluru, 560012, India}
\affil[3]{Department of Physics and Atmospheric Science, Dalhousie University, Halifax B3H 4R2, Nova Scotia, Canada}
\affil[*]{Email: \href{mailto:penghao.xiao@dal.ca}{penghao.xiao@dal.ca}; \href{mailto:saigautamg@iisc.ac.in}{saigautamg@iisc.ac.in}}

\date{}

\begin{document}

\maketitle

\begin{abstract}
Fast, and accurate prediction of ionic migration barriers ($E_m$) is crucial for designing next-generation battery materials that combine high energy density with facile ion transport. Given the computational costs associated with estimating $E_m$ using conventional density functional theory (DFT) based nudged elastic band (NEB) calculations, we benchmark the accuracy in $E_m$ and geometry predictions of five foundational machine learned interatomic potentials (MLIPs), which can potentially accelerate predictions of ionic transport. Specifically, we assess the accuracy of MACE-MP-0, Orb-v3, SevenNet, CHGNet, and M3GNet models, coupled with the NEB framework, against DFT-NEB-calculated $E_m$ across a diverse set of battery-relevant chemistries and structures. Notably, MACE-MP-0 and Orb-v3 exhibit the lowest mean absolute errors in $E_m$ predictions across the entire dataset and over data points that are not outliers, respectively. Importantly, Orb-v3 and SevenNet classify `good' versus `bad' ionic conductors with an accuracy of $>$82\%, based on a threshold $E_m$ of 500~meV, indicating their utility in high-throughput screening approaches. Notably, intermediate images generated by MACE-MP-0 and SevenNet provide better initial guesses relative to conventional interpolation techniques in $>$71\% of structures, offering a practical route to accelerate subsequent DFT-NEB relaxations. Finally, we observe that accurate $E_m$ predictions by MLIPs are not correlated with accurate (local) geometry predictions. Our work establishes the use-cases, accuracies, and limitations of foundational MLIPs in estimating $E_m$ and should serve as a base for accelerating the discovery of novel ionic conductors for batteries and beyond. 
\end{abstract}


\section{Introduction}
Developing next-generation batteries is essential for our transition into sustainable energy usage, given that the state-of-the-art lithium-ion batteries (LIBs), while already delivering excellent performance, are approaching their fundamental limits\cite{li201830, whittingham2014introduction}, necessitating the discovery of novel beyond-LIB materials and chemistries. One key materials property that governs battery performance is the ionic diffusivity ($D$) of the electroactive ion in electrodes and (solid) electrolytes. $D$ is exponentially influenced by the energy barrier each ion must overcome, commonly referred to as the migration barrier ($E_m$), to hop from its initial lattice site to a symmetrically equivalent final site\cite{park2010review,bachman2016inorganic,flores2022learning}.  Each ionic hop is often mediated through vacancies in the lattice with the ion overcoming transition state(s) along the hop. $D$ and $E_m$ are thus related via the Arrhenius expression, $D = D_0 \exp(-E_m/{k_B T})$\cite{vineyard1957frequency}.

Accordingly, materials with a low $E_m$ – in both electrodes and (solid) electrolytes exhibit higher ionic conductivity and enable faster charge/discharge rates\cite{van2020rechargeable}. In particular, emerging multivalent battery chemistries, such as Mg or Ca-based systems that promise higher volumetric energy densities\cite{canepa2017odyssey}, often suffer from poor rate performance\cite{orikasa2014high,liu2019multi,liang2020current,bayliss2019probing,gautam2016impact,wang2024challenges}. Some Na-ion cathodes such as marcite-Na(Mn/Fe)PO$_4$, phosphate alluaudite-Na$_x$MnFe$_2$(PO$_4$)$_3$ and sulfate sodium superionic conductors (NaSICONs) that offer lower costs compared to LIB cathodes also suffer from poor rate performance\cite{zhu2023first,deb2022critical}. Therefore, understanding and minimizing the $E_m$ in candidate materials is crucial for advancing the next-generation of high-performance batteries.

Experimental techniques such as quasi-elastic neutron scattering\cite{schwaighofer2025ionic}, electrochemical impedance spectroscopy\cite{Wang2021-oa}, nuclear magnetic resonance measurements\cite{clement2015manganese}, and galvanostatic intermittent titration techniques\cite{kang2021galvanostatic} are commonly employed to study ion dynamics in solids\cite{heitjans2006diffusion}. However, these methods often require access to large-scale facilities, and can exhibit chemistry or material-specific constraints/requirements, limiting their accessibility. As a result, computational approaches, particularly density functional theory (DFT\cite{hohenberg1964inhomogeneous,kohn1965self})-based nudged elastic band (NEB\cite{henkelman2000climbing}) calculations are commonly used for computationally estimating $E_m$ with reasonable precision\cite{devi2022effect}. While ab initio molecular dynamics (AIMD) simulations can also be used for estimating $E_m$, such simulations are computationally expensive since they require the sampling over large length and long time scales across different temperatures to provide reasonable $E_m$\cite{frenkel2023understanding,he2018statistical}. Moreover, AIMD simulations can be unreliable for systems exhibiting high $E_m$ (i.e., the `true negatives' among materials that can conduct ions) due to insufficient sampling of ion dynamics, resulting in DFT-NEB being the usual technique deployed for $E_m$ predictions.

Calculating $E_m$ using DFT-NEB requires an initial guess for the minimum energy path (MEP), which is typically constructed by linearly interpolating the coordinates of the initial and final configurations of the moving ion. Each `image' generated by linear interpolation is subsequently connected via an auxiliary spring force. Note that the initial interpolated guess is often far from the true MEP, increasing the computational expense of DFT-NEB calculations and making them prone to convergence difficulties\cite{devi2022effect}. Alternative approaches such as `ApproxNEB'\cite{rong2016efficient}, have been proposed to reduce computational intensity, with limited efficacy.

Recently, foundational machine learned interatomic potentials (MLIPs\cite{duval2023hitchhiker,unke2021machine,choi2025perspective}), also referred to as universal potentials, have emerged as a new paradigm in computational materials science. The foundational MLIPs, pre-trained on large and diverse datasets, can generalize to a wide range of downstream tasks\cite{choi2025perspective} and are transferable across different materials and property prediction tasks\cite{thiemann2024introduction,fu2022forces,jacobs2025practical,ju2025application}, unlike classical MLIPs or force-fields that are constrained by a specific chemistry or property. Thus, foundational MLIPs are attractive candidates for accelerating atomistic simulations, including NEB calculations, by potentially improving initial MEP guesses and reducing the need for extensive DFT-based refinement or optimization, which can enable high-throughput screening of materials based on their $E_m$. Indeed, a recent work by Kang et al.\cite{kang2025fasttrack} proposed an alternate to traditional DFT-based NEB calculations for $E_m$ estimations by using an MLIP for generating the potential energy surface on a spatial grid and extracting the MEP without the need for pre-defined NEB-images.

Several studies have benchmarked the performance of foundational MLIPs on diverse material properties\cite{focassio2024performance,loew2025universal,mannan2025evaluating,yu2024systematic}, but not on predicting $E_m$ in solids. For instance, the `Matbench discovery' platform\cite{riebesell2025framework} provides a standardized framework for ranking universal potentials, but does not yet evaluate their integration with NEB workflows for $E_m$ predictions. Other MLIP benchmarking studies include the work by Zhao et al.\cite{zhao2025harnessing} that evaluated the MLIPs on transition state search for chemical reactions involving molecules. Bihani et al.\cite{bihani2024egraffbench} benchmarked the performance of equivariant MLIPs on their generalizability to higher temperature simulations and unseen compositions, while Mannan et al. evaluated the performance of universal potentials against experimental measurements of elastic properties and structural accuracy among minerals\cite{mannan2025evaluating}. So far, there has been no benchmark of the performance of state-of-the-art universal potentials in predicting $E_m$ across a wide range of (battery) chemistries and materials, especially by integrating them with NEB workflows.

Here, we assess the performance of foundational MLIPs, namely, MACE-MP-0\cite{batatia2023foundation,batatia2025design}, SevenNet\cite{park2024scalable,kim2024data}, Orb-v3\cite{neumann2024orb,rhodes2025orb}, CHGNet\cite{deng2023chgnet}, and M3GNet\cite{chen2022universal} in predicting $E_m$ with NEB calculations. Using the dataset DFT-calculated $E_m$ compiled and curated by Devi et. al.\cite{devi2025literature,zenodo2025} that spans a wide range of materials and compositions, we benchmark the $E_m$ predictions of the foundational MLIPs against conventional DFT-NEB values at the generalized gradient approximation (GGA\cite{perdew1996generalized}) level of exchange-correlation accuracy for 574 migration paths. Additionally, we introduce a metric to assess the similarity of MLIP-NEB relaxed structures with the ground truth of DFT-NEB computed MEPs from our previous works\cite{devi2022effect,tekliye2022exploration,tekliye2024fluoride,deb2024exploration}. Finally, we examine the correlation between accuracies in $E_m$ and geometry predictions by the MLIPs considered.

Notably, we find that M3GNet and CHGNet tend to underestimate $E_m$ and exhibit a high degree of confidence in predicting low $E_m$ over a narrow range of possible $E_m$ values, while the other potentials exhibit no clear bias and deliver consistent accuracy over a wide range of $E_m$ values. Importantly, we observe that Orb-v3 and SevenNet classify systems as `good' ($E_m <$500~meV) or `bad' ionic conductors with $>$82\% accuracy. Performing an MLIP-NEB, using any of the potential considered, does result in improved interpolated paths representing the MEP in over 66\% of cases, indicating their utility in high-throughput screening workflows. Significantly, we find no evident correlation between the accuracy of $E_m$ and geometry predictions, with MLIPs yielding higher accuracy in $E_m$ predictions for systems with low $E_m$ values, while demonstrating better geometry predictions in systems with large $E_m$. We hope that our study establishes use-cases and quantifies the reliability of using foundational MLIPs in predicting $E_m$ over a diverse set of chemistries and crystal structures, which in turn should accelerate materials discovery for novel battery applications and beyond.

\newpage
\section{Methods}
\begin{figure}[h!]
\centering
\includegraphics[width=0.87\textwidth]{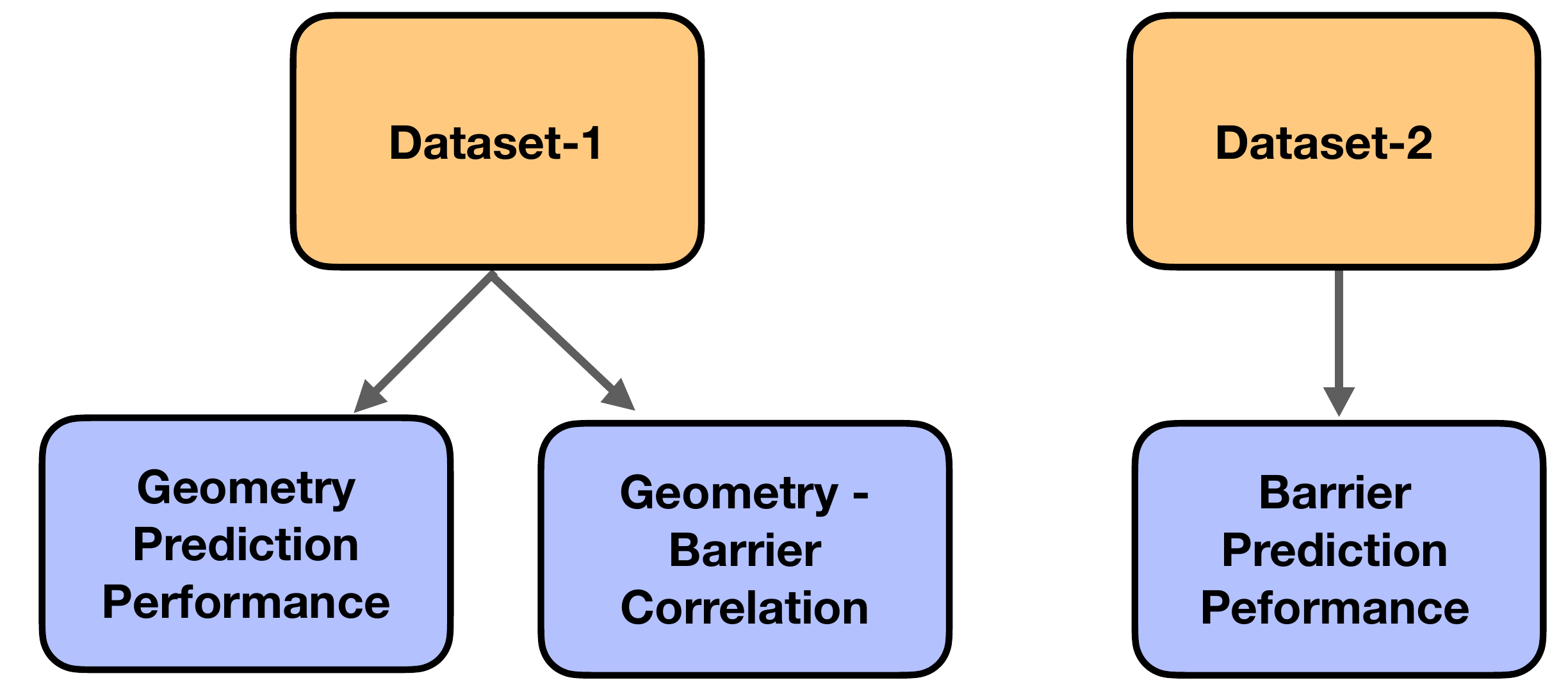}
\caption{Overview of the methodology, indicating the use of two subsets of the $E_m$ dataset that were created for examining geometry predictions, geometry-barrier correlations, and barrier predictions.}
\label{fig:fig0}
\end{figure}

\subsection{Datasets}
We utilize two distinct subsets of the $E_m$ dataset in this work. The smaller dataset, referred to as `Dataset-1', comprises 60 DFT-NEB calculations, including multiple possible migration pathways for select structures. Each datapoint consists of GGA-calculated $E_m$ and the relaxed structures of all intermediate images across a migration pathway. We constructed this dataset using the DFT-NEB results from our previous works\cite{devi2022effect,tekliye2022exploration,tekliye2024fluoride,deb2024exploration}, encompassing crystal structures explored primarily as battery materials, such as layered structures, Weberites, spinels, olivines, perovskites, and NaSICONs. The $E_m$ in Dataset-1 ranges from 0.06 eV to 2.88 eV.

The larger dataset, referred to as `Dataset-2', is a subset of a literature-derived collection of $E_m$\cite{devi2025literature}, which comprises of 621 DFT-calculated $E_m$ and the initial and final configurations for each migration pathway. Among the 621 datapoints, we excluded systems exhibiting $E_m >$2.5~eV, since such high $E_m$ values would not correspond to any tangible rate performance under battery operating conditions. We also excluded systems that presented significant convergence difficulties during NEB calculations using any of the foundational MLIPs considered ($\sim$10 datapoints), so that a fair and quantitative comparison can be made across the MLIPs. Thus, the final subset that forms our Dataset-2 consists of 574 systems. The systems comprising both datasets are compiled in our \href{https://github.com/sai-mat-group/mlips-migration-barriers}{GitHub} repository, while Dataset-2 is also available as a json file on \href{https://doi.org/10.5281/zenodo.17483476}{Zenodo}.

\subsection{Model Details}
We used publicly available universal potentials that have demonstrated good performance for bulk properties, are constructed on graph-based neural network (GNN) architectures, and are compatible with the atomic simulation environment (ASE\cite{larsen2017atomic}). We leveraged ASE calculators to integrate the MLIPs with the NEB implementation available within ASE. The specific MLIPs we used include $i$) the MACE-MP-0 `large' foundation model, trained on approximately 1.6 million materials project\cite{jain2013commentary} bulk-crystal relaxation trajectories (i.e., the `MPtrj' dataset\cite{deng2023materials}) with maximal message equivariance (L=2), $ii$) the SevenNet-MF-ompa model, which incorporates multifidelity learning with a core architecture based on the neural equivariant interatomic potential (NequIP\cite{batzner20223}) and is trained on MPtrj, `OMat24' trajectories\cite{barroso2024open}, and the subsampled Alexandria (sAlex\cite{schmidt2024improving}) datasets, $iii$) the Orb-v3-conservative-inf-omat that is trained on the MPtrj, OMat24 and Alexandria (Alex) dataset with learned forces being conservative by construction and effectively unlimited neighbor lists, $iv$) CHGNet v0.3.0, which is trained on MPtrj, and $v$) M3GNet MP-2021.2.8-EFS version, which is trained on MPtrj up to February 8, 2021. We used all models as provided, without any additional fine-tuning or hyperparameter optimizations. A summary of the specific details for each model is provided in Table~{\ref{tbl:mlips}}.

\begin{table*}[t]
\footnotesize
\setlength{\tabcolsep}{4pt}
\renewcommand{\arraystretch}{1.25}
\caption{Summary of the MLIPs used.}
\label{tbl:mlips}
\begin{tabularx}{\textwidth}{@{}>{\raggedright\arraybackslash}p{0.18\textwidth}
                                >{\raggedright\arraybackslash}p{0.28\textwidth}
                                Y@{}}
\hline
Model & Training data & Model type and key features \\
\hline
MACE-MP-0 & MPtrj dataset &
E(3)-equivariant GNN that captures many-body interactions \\
SevenNet-MF-ompa & MPtrj, OMat24, and sAlex &
Equivariant GNN incorporating multifidelity learning with efficient parallelization \\
Orb-v3 & MPtrj, OMat24, and Alex &
Roto-equivariance inducing regularized GNN with analytical energy gradients (conservative forces) and (effectively) infinite neighbors \\
CHGNet & MPtrj dataset &
GNN including magnetic moment inputs, thus incorporating information on atomic charges \\
M3GNet & MPtrj dataset &
Includes three-body interactions within its GNN \\
\hline
\end{tabularx}
\end{table*}

\subsection{NEB Calculations}
Typically, DFT-NEB calculations employ linear interpolation (LI) of atomic coordinates between the initial and final endpoints for generating the initial guess for the images. Whereas, the image dependent pair potential (IDPP) interpolation technique, developed by Smidstrup and coworkers\cite{smidstrup2014improved} utilizes a distance-matching objective to generate the initial guess for MEP. Given a preliminary benchmark of MACE-MP-0 based NEB calculations initialized with LI and IDPP interpolation, as detailed in Section~S1 and Figure~S1 of the supporting information (SI), we find IDPP interpolation to provide marginally better initial guesses for the eventual NEB calculations.

For NEB calculations of materials in Dataset-1 using all universal potentials, we generated seven intermediate images, mirroring the number used in the corresponding DFT-NEB calculations. The initial interpolated images were connected by a spring constant, $k$ = 5 eV/\AA$^2$, and we utilized the NEB implementation following the elastic band (EB,\cite{kolsbjerg2016automated}) method with full spring force, given our benchmarking with MACE-MP-0 (see Section~S1). We did not include the climbing image technique\cite{henkelman2000climbing} in any
of our MLIP-NEBs, as we did not see significantly different results
with or without climbing image in our previous work\cite{devi2022effect}. We deemed the NEB converged when the band forces fell below 0.05~eV/\AA, while using the Broyden–Fletcher–Goldfarb–Shanno optimizer\cite{broyden6convergence,fletcher1970new,goldfarb1970family,shanno1970conditioning}. In the case of Dataset-2, we employed only three intermediate images for all foundational MLIPs considered, to reduce computational costs. Also, we used the set of optimized NEB parameters that were used for calculations on Dataset-1 (i.e., $k$ = 5 eV/\AA$^2$, IDPP interpolation, and the EB method) for all calculations involving Dataset-2.

\subsection{Geometry Metrics}
To quantitatively assess the similarity of local geometries between structures, we introduce the geometric similarity classification metric, $\theta$ for a given image structure between the endpoints and an averaged geometric similarity score, $g$, across a migration path. $\theta$ evaluates whether the local geometry of an intermediate image obtained via a MLIP-NEB calculation is a better approximation of the reference structure (i.e., DFT-NEB relaxed) compared to the image generated using simple LI. Thus, $\theta$ is useful to determine whether intermediate images relaxed using MLIP-NEBs provide a superior initial guess for DFT-NEB calculations than typical LI, based on local geometric features. We define $\theta$ using the following steps:

\begin{enumerate}
    \item \textbf{Identify Nearest Neighbors and Pairwise Distances:}
    We identify the six nearest neighbors of the migrating ion using the Voronoi decomposition technique\cite{o1979proposed}, as implemented in the pymatgen package\cite{ong2013python}. Subsequently, we calculate all pairwise distances ($d$, using pymatgen) between the migrating ion ($c$) and its six neighbors ($i, j, k, l, m, n$), as well as among the neighbors themselves. The distances are calculated for structures relaxed/generated by DFT-NEB, MLIP-NEB, and LI. Thus, we compute, for any pair $\{x, y\} \subset \{i, j, k, l, m, n, c\}$ where $x \ne y$:
    $$d^{\text{DFT}}_{xy},\ d^{\text{MLIP}}_{xy},\ d^{\text{LI}}_{xy}$$

    \item \textbf{Calculate Absolute Errors in Pairwise Distances:}
    We compute the absolute difference between each pairwise distance in the MLIP-NEB relaxed structure and the LI structure with respect to the corresponding value in the DFT-NEB relaxed structure. These differences are stored in two 21-dimensional vectors:
    $$\boldsymbol{\Delta d^\text{MLIP}} = \left|d^{\text{DFT}}_{xy}-d^{\text{MLIP}}_{xy}\right|$$   $$\boldsymbol{\Delta d^\text{LI}} = \left|d^{\text{DFT}}_{xy}-d^{\text{LI}}_{xy}\right|$$
    for all $\{x, y\} \subset \{i, j, k, l, m, n, c\}$, where $x \ne y$.

    \item \textbf{Calculate Absolute Errors in Solid Angles:}
    Since two local geometries can have similar pairwise distances but differ in their angular orientations, we also consider the solid angles ($\Omega$, calculated using pymatgen) subtended by each face of the Voronoi polyhedra formed by the six nearest neighbors.
    $$\Omega^{\text{DFT}}_x, \quad \Omega^{\text{MLIP}}_x, \quad \Omega^{\text{LI}}_x, \quad \text{where} \quad x \in \{a, b, c, d, e, f\}$$
    In the above notation, two $\Omega$, say $\Omega^{\text{DFT}}, \Omega^{\text{MLIP}}$, having the same $x$ indicates that the polyhedral faces correspond to the same set of neighboring atoms. The absolute differences with the DFT-NEB relaxed structures are then calculated and stores as six-dimensional vectors:
    $$\boldsymbol{\Delta \Omega^\text{MLIP}} = \left|\Omega^{\text{DFT}}_{x}-\Omega^{\text{MLIP}}_{x}\right|$$   $$\boldsymbol{\Delta \Omega^\text{LI}} = \left|\Omega^{\text{DFT}}_{x}-\Omega^{\text{LI}}_{x}\right|$$
    for all $x \in \{a, b, c, d, e, f\}$.
\end{enumerate}

We expect the local geometry of an MLIP-NEB relaxed structure to be a poorer approximation of the DFT-NEB relaxed structure than the corresponding LI structure, if at least one of the following conditions is met: ($i$) one of the 21 pairwise distances or 6 solid angles of the MLIP-NEB relaxed structure deviates significantly more from the DFT-NEB geometry than the corresponding LI structure, or ($ii$) the average difference in pairwise distances or solid angles of the MLIP-NEB relaxed structure with the DFT-NEB reference is significantly higher compared to LI. To quantify these two conditions, we calculate $\delta$, which represents the maximum value among the differences in the mean and maximum errors of distances and angles between the MLIP-NEB and LI structures:

$$\delta = \max\Big( \overline{\boldsymbol{\Delta d^\text{MLIP}}} - \overline{\boldsymbol{\Delta d^\text{LI}}}, \quad \max(\boldsymbol{\Delta d^\text{MLIP}}) - \max(\boldsymbol{\Delta d^\text{LI}}),$$
$$\overline{\boldsymbol{\Delta \Omega^\text{MLIP}}} - \overline{\boldsymbol{\Delta \Omega^\text{LI}}}, \quad \max(\boldsymbol{\Delta \Omega^\text{MLIP}}) - \max(\boldsymbol{\Delta \Omega^\text{LI}}) \Big)$$

Here, $\overline{\boldsymbol{\Delta d}}$ and $\overline{\boldsymbol{\Delta \Omega}}$ represent the mean of the absolute errors in distances and solid angles, respectively. $\max(\boldsymbol{\Delta d})$ and $\max(\boldsymbol{\Delta \Omega})$ represent the maximum absolute errors.

Finally, the metric $\theta$ would classify the structure as:

\begin{equation}
\theta =
\begin{cases}
\text{Good}~\text{or}~1, & \delta < 0.01 \\[6pt]
\text{Comparable}~\text{or}~0, & 0.01 \leq \delta \leq 0.1 \\[6pt]
\text{Bad}~\text{or}~-1, & \delta > 0.1
\end{cases}
\label{eq:theta}
\end{equation}

Thus, \(\delta\) quantifies the difference between the deviations of the MLIP-NEB and LI structures with respect to the DFT-NEB reference based on key local geometric features. A smaller (ideally negative) \(\delta\) value signifies that the MLIP-NEB structure exhibits consistently lower errors, indicating it's a better approximation of the true DFT-NEB pathway. Conversely, larger (more positive) \(\delta\) suggests that LI performed as well or even better than the MLIP-NEB for at least one of the local geometric attributes. Therefore, we numerically represent the `good', `comparable' and `bad' structure as 1, 0 and -1 with $\theta$. Finally, for a given system containing $i$ intermediate images, we define \(g\) as,

\begin{equation*}
    g = \frac{\Sigma_i{\theta_i}}{\Sigma_ii}
\end{equation*}

In the case where all the $i$ image local geometries are better (worse) predicted by MLIP-NEB compared to LI, $g$ will take the value of 1 (-1). 

\section{Results}

\subsection{Barrier Prediction Performance}
\begin{figure}[h!]
\centering
\includegraphics[width=1.0\linewidth]{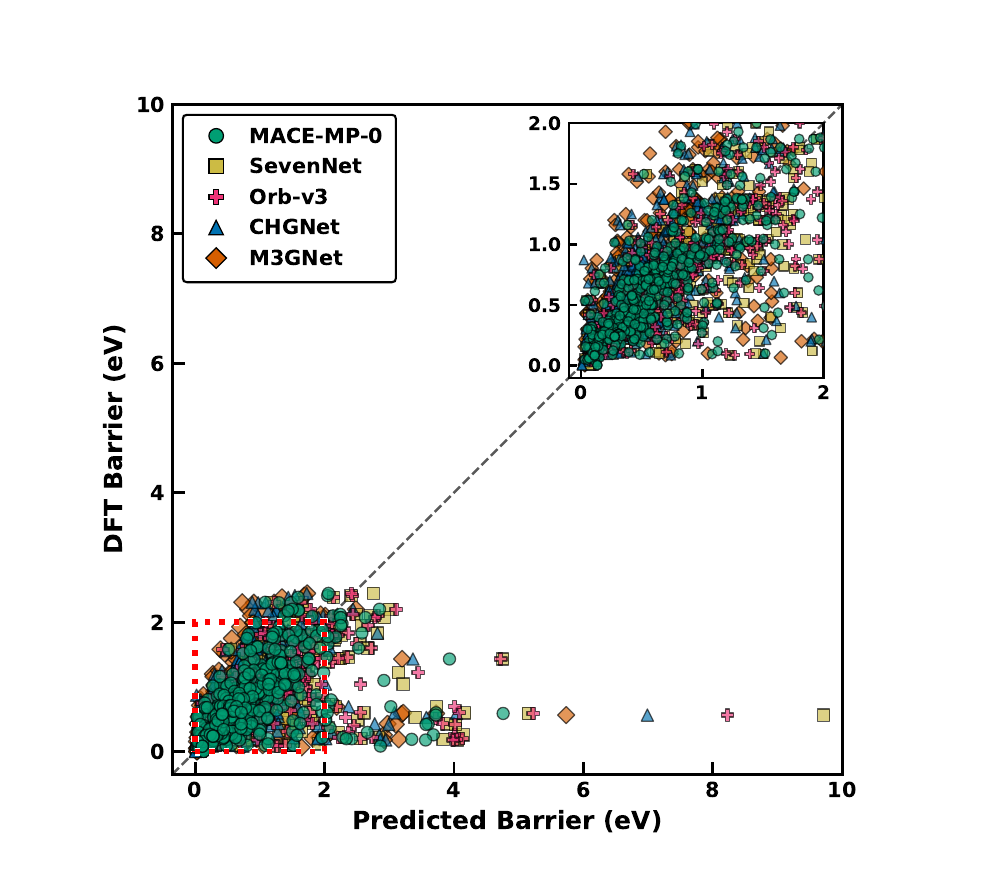}
\caption{Parity plot of migration barrier predicted by various MLIPs against DFT-NEB, with the dotted black line indicating the parity line. Inset shows the parity plot for a smaller range of DFT/predicted values (0-2~eV).}
\label{fig:fig2}
\end{figure}

Figure~{\ref{fig:fig2}} presents a comparison of $E_m$ predictions on Dataset-2 across different foundational MLIPs ($x$-axis) with their corresponding DFT-NEB calculated $E_m$ ($y$-axis). Green circles, yellow squares, pink pluses, blue triangles, and orange rhombuses represent $E_m$ predictions using MACE-MP-0, SevenNet, Orb-v3, CHGNet, and M3GNet, respectively. Individual parity plot for each MLIP considered is compiled in Figures~S2-S6 of the SI. Overall, MACE-MP-0 demonstrates the best performance with an MAE of 0.310~eV, while M3GNet records the highest MAE of 0.349~eV. The MAEs of Orb-v3, CHGNet, and SevenNet are 0.336, 0.343, and 0.344~eV, respectively. To provide a numerical context to the MAEs, DFT-NEB calculations typically carry a $\sim$60~meV error in their predictions, and a change of 60~meV in $E_m$ at 298~K corresponds to an order-of-magnitude change in $D$\cite{rong2015materials}. However, the calculation of MAEs is influenced by extreme outliers that affect all MLIPs.

To obtain a more representative picture of MLIP performance, we exclude 17 systems that act as common outliers across all MLIPs, with each outlier exhibiting absolute errors exceeding 1~eV. Notably, excluding the common outliers also reveals a similar performance hierarchy as with retaining the entire dataset: MACE-MP-0 emerges with the best MAE of 0.239~eV, followed closely by Orb-v3 with 0.245~eV. The remaining MLIPs, namely SevenNet, CHGNet, and M3GNet show MAEs of 0.251, 0.275, and 0.290~eV, respectively.

Besides accuracy, we analyze the distribution of datapoints relative to the ideal parity line to determine whether the MLIPs exhibit systematic prediction biases (i.e., under- or over-estimation of $E_m$). Interestingly, we observe MACE-MP-0, SevenNet, and Orb-v3 to demonstrate a relatively balanced prediction behavior with fairly symmetric distributions of under and over-estimated datapoints. Represented as (number of under-estimated datapoints, number of over-estimated datapoints) pairs, MACE-MP-0, SevenNet, and Orb-v3 exhibit distributions of (299,275), (244,330), and (242,332), respectively. In contrast, CHGNet and M3GNet show a bias toward under-estimating barriers, with under-estimated datapoints accounting for 73.1\% and 78.2\% of all predictions, respectively. Represented as (under-estimated, over-estimated) pairs, CHGNet and M3GNet exhibit distributions of (420,154) and (449,125), respectively.

To further understand individual MLIP capabilities, we examined each potential's performance after excluding the outliers specific to each potential (i.e., systems with absolute errors $>$ 1~eV as predicted by a given potential) to gain insight into the `best-case' scenario of $E_m$ predictions. Notably, despite having 37 outliers, Orb-v3 achieves the lowest MAE of 0.198~eV on its remaining (non-outlier) predictions. With 35 outliers, MACE-MP-0 is a close second with an MAE of 0.202~eV, while SevenNet, with 37 outliers, displays an MAE of 0.203~eV. CHGNet and M3GNet show higher MAE values of 0.248~eV and 0.257~eV with 31 and 36 outliers, respectively. Thus, we find that Orb-v3 can achieve higher accuracies on systems that it describes well while MACE-MP-0 achieves a better balance of both low errors and fewer outliers compared to other MLIPs.

\subsection{Predictions Over Different Barrier Ranges}
\begin{figure}[h!]
\centering
\includegraphics[width=1.0\linewidth]{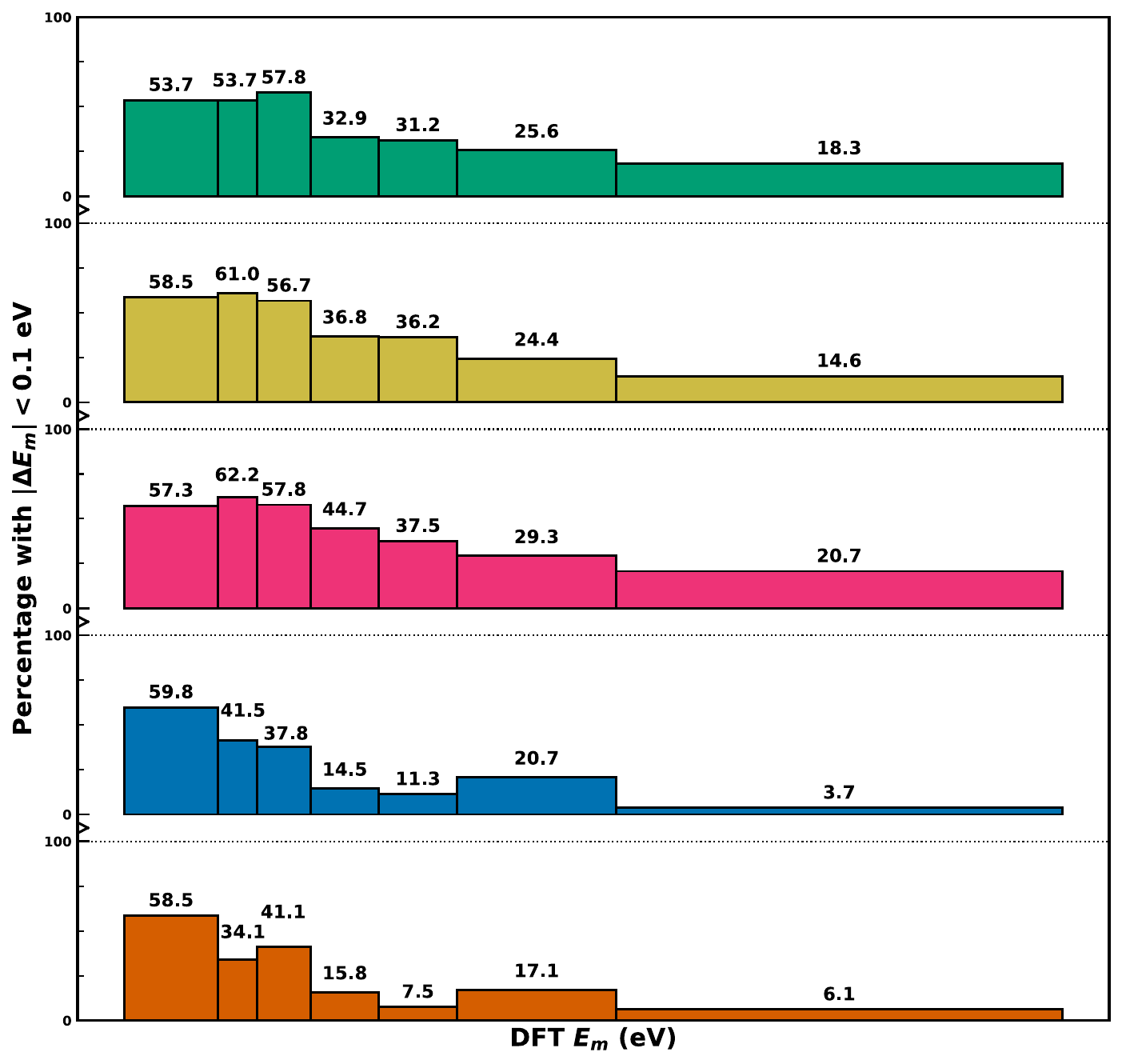}
\caption{Barrier prediction performance of various MLIPs across different DFT-calculated $E_m$ ranges. The dotted line and kink denote a change in the models, which are, from top to bottom: MACE-MP-0, SevenNet, Orb-v3, CHGNet, and M3GNet. Each bin contains an equal number of data points with the width corresponding to the range of DFT-calculated $E_m$ within the bin. The height of each bin (as indicated by the numerical annotation on each bin) within each model represents the percentage of data points whose $E_m$ values are predicted within an absolute error of 0.1~eV.}
\label{fig:fig3}
\end{figure}
To understand the performance of MLIPs across various ranges of $E_m$, we divided the 574 datapoints into seven equal-sized distributions based on their DFT-NEB $E_m$ values, as illustrated in Figure~{\ref{fig:fig3}}. While the $x$-axis in Figure~{\ref{fig:fig3}} represents DFT-NEB $E_m$ ranges, with bar widths indicating the span of $E_m$ values within each bin, the $y$-axis shows the percentage of predictions within each bin that achieve absolute errors $<$0.1~eV (signifying an ``acceptable'' degree of accuracy). The exact DFT barrier range of the data points present in each bin can be found in Table~S1 of the SI. Trends in Figure~{\ref{fig:fig3}} indicate that all MLIPs struggle with high $E_m$ predictions, with only small percentage of systems exhibiting the acceptable accuracy in the highest barrier range ($\sim$1.31-2.50~eV). Specifically, the percentage of predictions with acceptable accuracy in the highest $E_m$ range are 20.7\% for Orb-v3, 18.3\% for MACE-MP-0, 14.6\% for SevenNet, 6.1\% for M3GNet, and 3.7\% for CHGNet. 

Importantly, we identify a ``sweet spot'' of $E_m$ values where all MLIPs perform reasonably well. For example, in the low-barrier range ($\sim$0.0025-0.25~eV), more than 50\% of predictions achieve acceptable accuracy across all MLIPs. Within this range, CHGNet shows the highest success rate (59.8\%), followed by M3GNet and SevenNet (both 58.5\%), while MACE-MP-0 and Orb-v3 achieve 53.7\% and 57.3\%, respectively. Additionally, Orb-v3 and SevenNet achieve their best performance (i.e., highest fraction of predicted datapoints with acceptable accuracy) in the 0.25-0.36~eV range, achieving 62.2\% and 61\% acceptable predictions, respectively. MACE-MP-0 performs best in the slightly higher 0.36-0.50~eV range with 57.8\% accuracy. Meanwhile, CHGNet and M3GNet perform best in the lowest $E_m$ range ($\sim$0.0025-0.25~eV) with 59.8\% and 58.5\% accuracy, respectively.

While all MLIPs show declining accuracy with increasing $E_m$, Orb-v3 exhibits the slowest degradation, maintaining better performance across a broader range of $E_m$ values compared to other potentials. Thus, we find that `simpler' graph models such as CHGNet and M3GNet demonstrate superior performance for materials with intrinsically low $E_m$ values but lack consistency in their predictions over a wider range of $E_m$. On the other hand, increasing complexity among the graph models, such as in Orb-v3 or MACE-MP-0 allows for a more robust performance across a wide range of $E_m$ values while sacrificing `peak' performance for materials with low $E_m$, making them better suited for $E_m$ predictions in novel materials. This variation in the performance of `simple' and `complex' MLIPs also reveals the general trade-off between building specialized and generalized models in the field of machine learning.

\subsection{Barrier Classification Performance}
\begin{figure}[h!]
\centering
\includegraphics[width=1.0\linewidth]{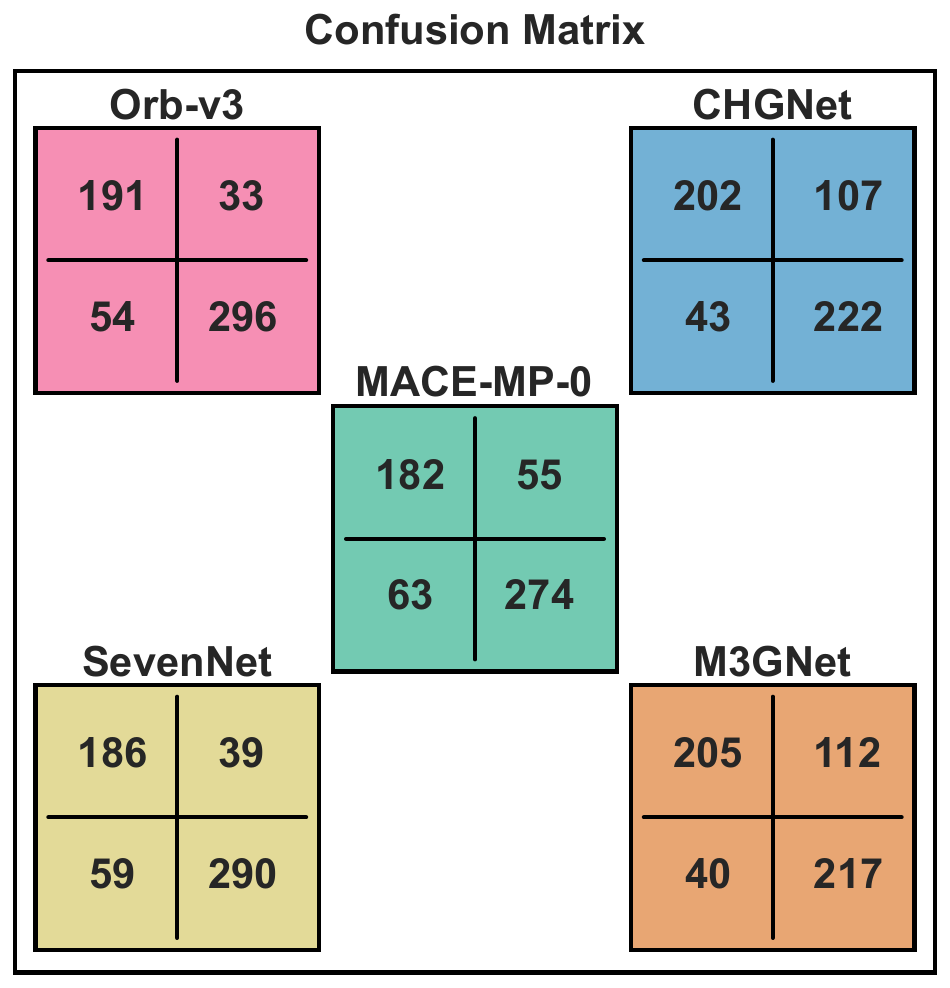}
\caption{Confusion matrices for barrier prediction across different models. Each matrix corresponds to a specific model and is structured such that the upper-left, upper-right, lower-left, and lower-right cells represent the counts of true positives (TP), false positives (FP), false negatives (FN), and true negatives (TN), respectively. A prediction is considered a TP (TN) if both the DFT-computed and model-predicted $E_m$ are less than (greater than or equal to) 500~meV.}
\label{fig:fig4}
\end{figure}
To quantify the ability of the MLIPs considered to classify a material as a `good' versus a `bad' ionic conductor, which can be used for high-throughput identification of promising candidates, we present the confusion matrices for all MLIPs in Figure~{\ref{fig:fig4}}. Each potential in Figure~{\ref{fig:fig4}} is represented using a distinct color, such as green (MACE-MP-0), yellow (SevenNet), pink (Orb-v3), blue (CHGNet), and orange (M3GNet). For the classification task, we use a threshold $E_m$ of 500~meV, i.e., materials that exhibit a calculated/predicted $E_m <$500~meV are labeled good ionic conductors, while materials that show higher values of $E_m$ are labeled bad ionic conductors. Within each confusion matrix, the true positive (TP), the true negative (TN), the false positive (FP) and the false negative (FN) numbers are listed on the top left, bottom right, top right, and bottom left cells, respectively. 

From Figure~{\ref{fig:fig4}}, we observe that Orb-v3 achieves the highest combined number of TP and TN, correctly classifying 487 out of 574 systems (i.e., an accuracy of 84.84\%). In comparison, M3GNet yields the lowest TP+TN count of 422 systems (73.52\%). SevenNet, MACE-MP-0 and CHGNet correctly classify 476 (82.93\%), 456 (79.44\%), and 424 (73.87\%) systems, respectively. These results highlight Orb-v3 as the most reliable model for distinguishing good and poor ionic conductors, followed closely by SevenNet (accuracy $>$ 80\%), making both reliable for high-throughput classification tasks.

\subsection{Geometry Prediction Performance}
\begin{figure}[h!]
\centering
\includegraphics[width=1.0\linewidth]{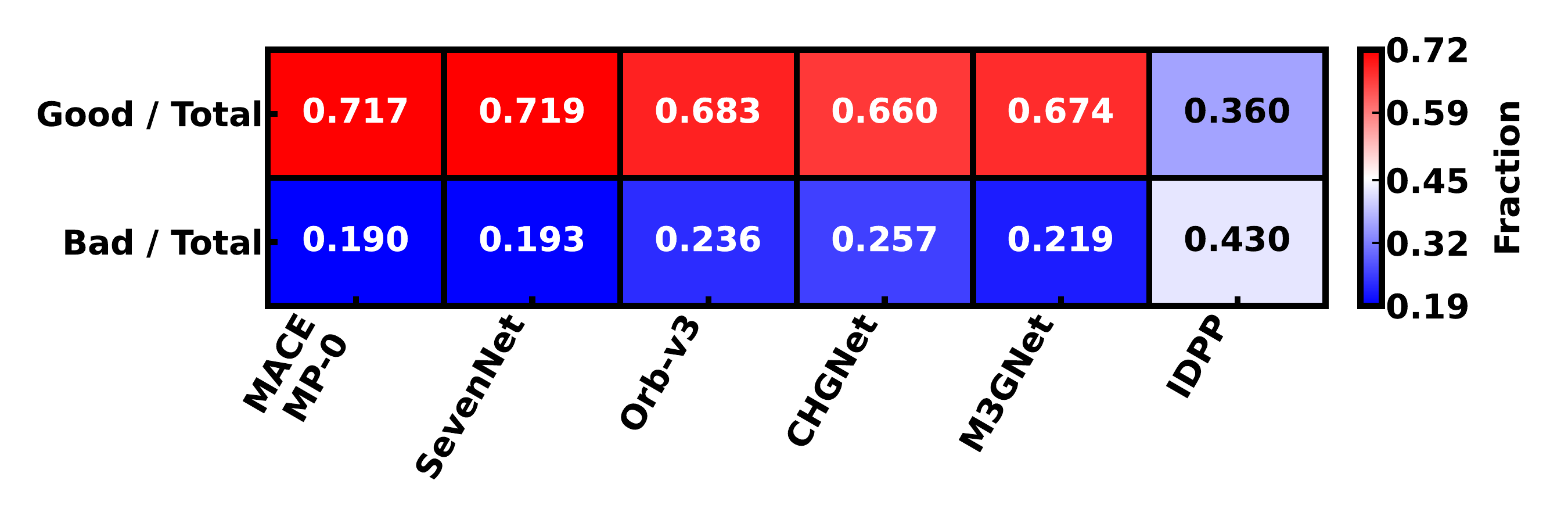}
\caption{Performance of MLIPs on local geometry prediction. Each entry in the heatmap represents a performance fraction for a given MLIP with the last column corresponding to IDPP. The top (bottom) row shows the fraction of structures classified as `good' (`bad') to the total number of structures. The heatmap color bar varies from red (high fractions) to blue (low fractions).}
\label{fig:fig5}
\end{figure}
It is important for any MLIP to not only get the $E_m$ correct but also get the underlying geometries that constitute the MEP (and hence yield the $E_m$ value) correct for the MLIP to be truly accurate. Thus, we quantify the performance of the MLIPs considered in their predictions of local geometry of the intermediate image structures in Dataset-1 (using $\theta$ of Equation~{\ref{eq:theta}}) as a heatmap in Figure~{\ref{fig:fig5}}. Note that we performed an NEB calculation with EB, $k = 5$~eV/\AA{}$^2$, and IDPP interpolation with each MLIP and for each material in Dataset-1 to generate the statistics displayed in Figure~{\ref{fig:fig5}}. For each MLIP ($x$-axis), we denote the fraction of structures with `good' (top row) and `bad' (bottom) local geometries in Figure~{\ref{fig:fig5}}. Ideally, the MLIPs should exhibit a high (low) fraction of structures with good (bad) local geometry. Since IDPP (without any subsequent MLIP-based relaxation) also behaves like a potential for generating a guess for the MEP, we include IDPP's statistics in Figure~{\ref{fig:fig5}}.

Among all the MLIPs, SevenNet exhibits the highest fraction of good geometries (0.719), indicating that it frequently generates accurate local geometries. On the other hand, MACE-MP-0 exhibits the lowest fraction of bad geometries (0.190), indicating that it frequently avoids generating inaccurate structures. The difference between the fraction of good and bad geometry predictions for both MACE-MP-0 and SevenNet are similar (0.527 and 0.526, respectively), indicating that both models perform equally well in generating good local geometries. 

Other MLIPs show poorer geometry predictions, with Orb-v3, M3GNet, and CHGNet displaying good (bad) fractions of 0.683 (0.236), 0.674 (0.219), and 0.660 (0.257), respectively, with CHGNet showing the smallest difference between the good and bad geometry fractions (0.403). Thus, MACE-MP-0 and SevenNet show significantly better local geometry predictions upon relaxation with NEB compared to Orb-v3, M3GNet and CHGNet, while all MLIPs provide better initial guesses to the MEP than LI in at least 66\% of structures (i.e., intermediate images). Also, we note that IDPP generated structures are statistically much farther from DFT than MLIPs, with LI being better than IDPP in 43\% of the cases. Given our definition of \(\theta\)  and the specific systems present in Dataset-1, we find that IDPP does not make a significant difference in enhancing the initial guess for the MEP as compared to LI across all MLIPs.

\subsection{Geometry-Barrier Correlation}
\begin{figure}[h!]
\centering
\includegraphics[width=1.0\linewidth]{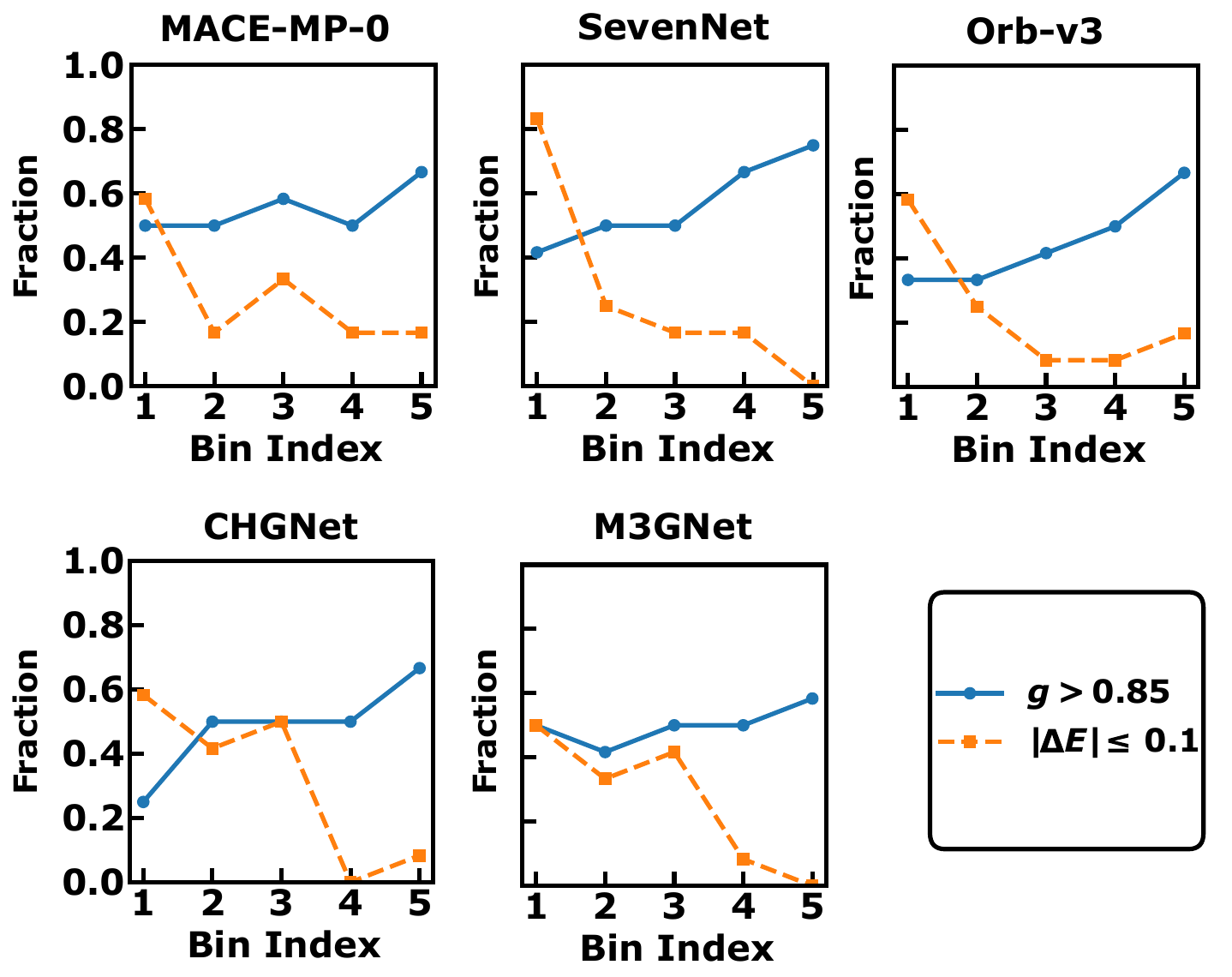}
\caption{Correlation between $E_m$ and local geometry prediction. Blue circles correspond to the fraction of data points within a given bin that has $g >$0.85, while the orange rectangles represent the fraction of data points in the same bin which have an absolute error in the $E_m$ prediction, $\Delta E \leq$0.1~eV.}
\label{fig:fig6}
\end{figure}
To investigate whether there is a correlation between geometry and $E_m$ prediction performance, i.e., whether a precise $E_m$ prediction by a given MLIP is due to its precise geometry prediction, we divide Dataset-1 into five bins based on their DFT-NEB $E_m$, ensuring that each bin contains an equal number of data points. Bins with indices 1, 2, 3, 4, and 5 correspond to $E_m$ ranges of [0.058,0.64], (0.64,0.97], (0.97,1.24], (1.24,1.81], and (1.81,2.88] eV, respectively. Note that the number of bins in Figure~{\ref{fig:fig6}} is different from that of Figure~{\ref{fig:fig3}} since the datasets considered for both figures are different. Within each bin, we estimate the fraction of migration paths with `good' local geometry ($g >$0.85, blue points) and low absolute errors in $E_m$ ($\Delta E\leq$0.1~eV, orange points), and plot the statistics for each MLIP in Figure~{\ref{fig:fig6}}. Note that $g >$ 0.85 signifies cases where the predicted geometry is better than unrelaxed LI for at least six out of the seven intermediate images.

Overall, Figure~{\ref{fig:fig6}} reveals the absence of any positive correlation between barrier and geometry prediction performance, and more strikingly, an inverse relationship. For example, all models perform poorly in predicting high $E_m$ (bin 5), which is consistent with our observations in Figure~{\ref{fig:fig4}}. However, all models also achieve their best geometry predictions for bin 5. In other words, the best geometry predictions are coincident with the worst $E_m$ predictions. The geometry prediction success rates within bin-5 are 66.7\%, 75.0\%, 66.7\%, 66.7\%, and 58.3\%  for MACE-MP-0, SevenNet, Orb-v3, CHGNet, and M3GNet, respectively, and the corresponding $E_m$ prediction success rates (i.e., $\Delta E\leq$0.1~eV) are 16.7\%, 0\%, 16.7\%, 8.3\%, and 0\%, respectively.

To further assess geometry-barrier correlation, we examine instances where MLIPs perform well in both metrics. Note that, we term a model to exhibit a `good performance in both metrics' if both the fractions in a given bin are $>$ 0.5. Only two potentials show this good performance, and only in a single bin (Bin-1), namely, MACE-MP-0 with a success rate of 58.3\% in $E_m$ prediction and 50.0\% in geometry prediction, and M3GNet with a 50.0\% success rate for both metrics. SevenNet, Orb-v3, and CHGNet do not achieve this good performance in any bin. Moreover, we find no consistent pattern across all bins and all MLIPs and no instances where good $E_m$ predictions coincide with good geometry prediction. Instead, the data suggests that these two performance metrics are largely independent, and that a good $E_m$ prediction does not necessarily arise from a good local geometry prediction (and vice-versa).

\section{Discussion}
Given the critical role of $E_m$ in battery materials design and the high computational costs associated with DFT-NEB calculations, we have evaluated the performance of foundational MLIPs including MACE-MP-0, SevenNet, Orb-v3, CHGNet, and M3GNet, for $E_m$ predictions upon integration with the NEB framework over two data subsets containing $E_m$ and structural data (Figure~\ref{fig:fig0}). Specifically, we investigated $i$) the ability of MLIPs to predict $E_m$ accurately, $ii$) the likelihood of generating MLIP-NEB-relaxed image geometries that are close to the ground truth (DFT-NEB), and $iii$) whether any correlation exists between the accuracy in $E_m$ prediction and geometry relaxation.

Analyzing $E_m$ predictions across the entire Dataset-2, we find that MACE-MP-0 exhibits the lowest MAE (Figure~{\ref{fig:fig2}}), followed in order by Orb-v3, CHGNet, SevenNet, and M3GNet. On excluding outliers that are common to all models, we observe SevenNet to exhibit a slightly lower MAE than CHGNet, with the rest of the performance order being the same. Interestingly, when assessing each model independently after removing their respective outliers, Orb-v3 demonstrates the best MAE of 0.198~eV, marginally outperforming MACE-MP0 (0.202~eV), with the other models exhibiting larger errors (0.203-0.257~eV). Thus, Orb-v3 provides the best prediction errors for $E_m$, among MLIPs considered, in systems with a robust description of the corresponding potential energy surface. 

Notably, during endpoint relaxations for Orb-v3, 153 systems failed to converge within the threshold forces over 1000 optimization steps despite attempting multiple optimization algorithms. However, to maintain consistency with the other models in the study, we did not modify the obtained results and included the Orb-v3 results as is. Nevertheless, more robust relaxation strategies with extended optimization steps may enable Orb-v3 to achieve better accuracies in $E_m$ predictions, potentially surpassing the other MLIPs considered across the entire Dataset-2. 

We observe that M3GNet and CHGNet exhibit a systematic bias toward underestimating $E_m$, whereas MACE-MP-0, SevenNet, and Orb-v3 do not display such a tendency. A more granular analysis (Figure~{\ref{fig:fig3}}) reveals that all models struggle with accurately predicting high $E_m$ values. Among them, Orb-v3 shows a relatively slow decay in prediction accuracy as the $E_m$ value increases. Interestingly, the simpler models CHGNet and M3GNet outperform their more complex counterparts within a very narrow range of low $E_m$ but exhibit a rapid decline in performance as the range expands.

Using a threshold $E_m$ of 500 meV to categorize structures as `good' or `bad' conductors of ions (Figure~{\ref{fig:fig4}}), we find that all MLIPs are able to identify good conductors with reasonable accuracy ($>$73\%). Orb-v3 and SevenNet display the highest accuracies in classifying good (or bad) conductors, with $\sim$85\% and $\sim$83\% accuracy, respectively, making them highly suitable for high-throughput screening of candidate battery materials. 

Our study on Dataset-1 indicates that MLIP-NEB relaxations tend to produce image geometries that are as close as (or closer to) DFT-NEB structures than those obtained through simple LI or IDPP interpolation in the majority ($\sim$66\%, Figure~{\ref{fig:fig5}}) of cases. Among the considered models, MACE-MP-0 and SevenNet stand out in geometry predictions, relaxing to geometries that are worse than LI or IDPP ones in only 19\% of migration paths, suggesting that employing MACE-MP-0 or SevenNet NEB-relaxed images as initial guesses for DFT-NEB calculations could significantly accelerate convergence and reduce computational costs. Note that although our metric, $\theta$ (see Equation~{\ref{eq:theta}}), captures critical local geometric features, it can be improved further to decisively quantify local structural similarity.

Finally, when simultaneously evaluating the likelihood of accurate barrier prediction and better geometry initialization (Figure~{\ref{fig:fig6}}), we observe no evident correlation between the two among all MLIPs considered. Thus, we find that accurate barrier predictions do not necessarily imply better geometry predictions, and vice versa. One possible explanation for this counterintuitive trend is that for systems with low $E_m$, the potential energy surfaces are likely `flat' with variations in local geometries, indicating that even large errors in local bond distances or local bond angles made by the MLIPs do not significantly change the predicted $E_m$, thus leading to accurate $E_m$ even with inaccurate geometries. On the other hand, for systems with large $E_m$, the potential energy surfaces should exhibit `deep' minima associated with the `stable' sites occupied by the migrating ion, signifying that even small errors in predicting local bond distances or angles by the MLIPs can cause large errors in the $E_m$ predictions, thus resulting in inaccurate $E_m$ even with mostly accurate geometries.

\section{Conclusion}
Given the importance of accurate and swift predictions of $E_m$ in materials for battery applications (and beyond), we systematically evaluated five foundational MLIPs (MACE-MP-0, SevenNet, Orb-v3, CHGNet, and M3GNet) via integration with the NEB framework across a diverse set of chemistries and materials relevant for batteries. We benchmarked the barrier prediction accuracy against DFT-NEB calculated values for all MLIPs, and found MACE-MP-0 to achieve the lowest MAE (0.310~eV), while Orb-v3 demonstrated significantly better performance (MAE of 0.198~eV) when evaluated on data points that were not outliers. Further, we assessed the capability of the MLIPs to classify materials as good ($E_m<$ 500 meV) or bad ($E_m \ge$ 500 meV) ionic diffusers, and observed Orb-v3 and SevenNet to accurately classify $>$82\% of migration paths, making them suitable for high-throughput screening applications. Based on our novel geometric similarity metric, we demonstrated that MLIP-NEB relaxations produce image structures that are closer to DFT-NEB calculated references than LI in over 66\% of cases. Finally, we discovered no direct correlation between barrier prediction accuracy and the similarity of the MLIP-NEB relaxed geometry to the DFT-NEB reference. We hope that our work establishes use-cases, accuracies, and limitations in using foundational MLIPs for predicting and/or accelerating $E_m$ calculations, which should result in better discovery of novel ionic conductors with applications as electrodes and (solid) electrolytes in batteries and other related technologies.

\section{Data availability}
Both datasets and associated python scripts used in this work are compiled and available for free at our \href{https://github.com/sai-mat-group/mlips-migration-barriers}{GitHub} repository. Dataset-2 is available as a json file on \href{https://doi.org/10.5281/zenodo.17483476}{Zenodo}.

\section{Acknowledgements}
G.S.G. acknowledges financial support from the Science and Engineering Research Board (SERB) of the Department of
Science and Technology, Government of India, under sanction number IPA/2021/000007. A.K.B. thanks the Ministry of Human Resource Development, Government of India, for financial assistance. 
The authors gratefully acknowledge the super-computing facility offered by ACENET and the Digital Research Alliance of Canada. The authors acknowledge the computational resources of the super computer ‘PARAM Pravega’ provided by the super computer education and research centre (SERC) at IISc. The authors also thank the Jülich Supercomputing Centre (at Forschungszentrum Jülich), Germany for the use of the ‘JURECA’ supercomputer, under projects `hpc-prf-emdft' and `hpc-prf-desal'. 


\section{Competing Interests}
The authors declare no competing financial or non-financial interests.


\bibliography{rsc} 
\bibliographystyle{unsrt}

\end{document}